\documentclass{elsart}
\usepackage{amsmath}
\usepackage{amssymb}
\usepackage{graphicx}

\def\bx#1{\leavevmode\thinspace\hbox{\vrule\vtop{\vbox{\hrule\kern1pt
        \hbox{\vphantom{\tt/}\thinspace{\bf#1}\thinspace}}
      \kern1pt\hrule}\vrule}\thinspace}

\def\be{\begin{equation}}
\def\ee{\end{equation}}
\def\ba{\begin{eqnarray}}
\def\ea{\end{eqnarray}}

\begin{document}
\begin{frontmatter}

\title{Redshift and Shear Calibration: Impact on Cosmic Shear Studies and Survey Design}

\author{L. Van Waerbeke$^1$, M. White$^{2,3}$, H. Hoekstra$^4$, C. Heymans$^1$}

\address{$^1$Department of Physics and Astronomy, University of British Columbia, Vancouver, BC V6T 1Z1, Canada.}
\address{$^2$Departments of Physics and Astronomy, University of California\\
Berkeley, CA 94720, USA.}
\address{$^3$Lawrence Berkeley National Laboratory, 1 Cyclotron Road\\
Berkeley, CA 94720, USA.}
\address{$^4$ Department of Physics and Astronomy, University of Victoria, Victoria, BC V8P 5C2, Canada.}

\begin{abstract}
The cosmological interpretation of weak lensing by large-scale
structures requires knowledge of the redshift distribution of the
source galaxies. Current lensing surveys are often calibrated
using external redshift samples which span a significantly smaller
sky area in comparison to the lensing survey, and are thus subject
to sample variance. Some future lensing surveys are expected to be
calibrated in the same way, in particular the fainter galaxy
populations where the entire color coverage, and hence photometric
redshift estimate, could be challenging to obtain. With N-body
simulations, we study the impact of this sample variance on cosmic
shear analysis and show that, to first approximation, it behaves
like a shear calibration error $1\pm\epsilon$. Using the Hubble
Deep Field as a redshift calibration survey could therefore be a
problem for current lensing surveys. We discuss the impact of the
redshift distribution sampling error and a shear calibration error
on the design of future lensing surveys, and find that a lensing
survey of area $\Theta$ square degrees and limiting magnitude
$m_{\rm lim}$, has a minimum shear and redshift calibration
accuracy requirements given by $\epsilon = \epsilon_0
10^{\beta(m_{\rm lim}-24.5)}\left(\Theta/ 200\right)^{-1/2}$.
Above that limit, lensing surveys would not reach their full
potential. Using the galaxy number counts from the Hubble
Ultra-Deep Field, we find $(\epsilon_0,\beta)=(0.015,-0.18)$ and
$(\epsilon_0,\beta)=(0.011,-0.23)$ for ground and space based
surveys respectively. Lensing surveys with no or limited redshift
information and/or poor shear calibration accuracy will loose
their potential to analyse the cosmic shear signal in the
sub-degree angular scales, and therefore complete photometric
redshift coverage should be a top priority for future lensing
surveys.
\end{abstract}


\end{frontmatter}

\section{Introduction}

Weak gravitational lensing by large-scale structure probes the
matter distribution in the nearby Universe, regardless of where
the `light' baryonic matter is with respect to dark matter. While
to first order the calculation of the deflection of light by
large-scale structure is easy, the details of the propagation
depend upon the 3-dimensional distribution of matter. In a
modern survey the same mass can therefore act as both lens and source
\cite{LensReview}. In order to infer cosmology from lensing
measurements it is thus crucial to know the source redshift
distribution. To first approximation, the lensing effect depends
on the mean source redshift, but in reality it depends on the full
distribution function. Cosmic shear measurements to date
(see \cite{WaeMel,HoeYeeGla,STEP1} for a recent compilation)
assume a mean source redshift calibrated, at least in part,
from an external spectroscopic or photometric redshift sample. The current
treatment is to derive a direct translation of magnitude into
redshift from the calibration sample.
The problem with redshift samples is that they cover a very small
area of the sky in comparison to the lensing surveys. Therefore
there is a risk that the calibration sample is {\it too small\/}
and might be subject to significant sample variance. The purpose
of the this paper is to study the impact of the source redshift
distribution sample variance of calibration samples on weak
lensing analysis, and in particular on cosmological parameter
estimation. Although most future lensing surveys plan to get full
photometric redshift coverage, and are therefore potentially
unaffected by this error (only if the photometric redshifts are
unbiased), some may not\footnote{For instance the current DUNE
baseline plans to have a single band imaged from space and partial
color follow-up from the ground \cite{ref}}. Our analysis could
then be used to estimate realistic redshift sampling errors and as
a reference to aid the design of an optimal redshift calibration
survey. Another motivation for this work is also to address the
choice of the CFHTLS-WIDE to postpone the accurate measurement of
photometric redshifts to the end of the survey, and study to which
extent using external redshift calibration fields will affect the
parameter accuracy.

This work complements recent analyses along the same theme. In
\cite{HTBJ}, the authors investigate the effect photometric
redshift errors have on the redshift distribution. They describe
the error on $n(z)$ as a set of polynomials and calculate the
corresponding error on the measured cosmological parameters. They
do not address the redshift sampling variance issue, however. A
preliminary investigation of the effect of source clustering in
the redshift distribution was performed in \cite{Ish} (see also
\cite{HTBJ,MHH} for related work). In \cite{Ish} the authors
assumed that the source distribution follows a 3-dimensional
Gaussian distribution, and they made no distinction between
different types of source galaxies. They estimated how many
galaxies needed to be targeted for spectroscopy in order to get a
good estimate of the redshift distribution that reduces the sampling
variance to an acceptable value. The two limitations of their work
are 1) galaxies are in fact subject to non-Gaussian clustering,
which increases the effect of sampling variance and 2) galaxies
come with a variety of masses, magnitude and shapes that are
correlated with their redshift. Their work was therefore a study
of a homogeneous incompleteness of redshift information in a
lensing survey assuming Gaussian statistics.  It was also limited to large
scales only ($l<3000$).

In this paper we extend these previous analyses by including
realistic source clustering of the galaxy population. We also
derive more general requirements regarding the redshift
calibration sample for different observing strategies. In our
work, the redshift calibration sample could be a distinct survey
from the lensing survey, as is the case for the VIRMOS
\cite{LVW05}, RCS \cite{HH02}, CFHTLS \cite{HH06,ES06}, WHT
\cite{WHT}, Groth strip \cite{Groth}, MDS \cite{MDS} and STIS
\cite{STIS} surveys which all used the small field-of-view Hubble
Deep Fields (HDF) \cite{HDF} for example, or it could also be part
(or all) of the lensing survey, as is the case for the COMBO-17
\cite{COMBO-17} and GEMS \cite{Heymans05} surveys. Significant
efforts are under way in order to improve our knowledge of the
galaxy redshift distribution (VVDS \cite{VVDS}, DEEP2 \cite{DEEP2} and
zCOSMOS\footnote{zCOSMOS: www.exp-astro.phys.ethz.ch/zCOSMOS/}).
We therefore also address to which extent these spectroscopic surveys
can be used to calibrate on-going and future lensing surveys with
only partial photometric redshift coverage. We establish the
limits to which the HDF can be safely used, and in particular we
quantify the covariance of the redshift distribution for different
redshift calibration survey areas.

The next Section introduces the notation and relevant quantities
used in this work.  It also describes the construction of the mock
galaxy catalogues used to model the source redshift sample
variance. Section \ref{sec:results} discusses how the sample
variance affects the cosmological parameters.  In Section 4 we
discuss how the future design of weak lensing surveys should take calibration
issues into account and Section 5 discusses the impact of calibration issues
on previous weak
lensing measurements.  We conclude in Section \ref{sec:conclusions}.

\section{Method}  \label{sec:description}

\subsection{Background}

There are numerous ways of constraining cosmological models from weak lensing
data, but the most common uses a 2-point statistic such as the shear
correlation
function or shear variance smoothed on a range of scales.  The smoothed shear
variance $\langle\gamma^2\rangle$
is related to the power spectrum (or Fourier transform of the
shear correlation function) by \cite{LensReview}
\begin{equation}
  \langle\gamma^2\rangle = {2\over \pi\theta^2}
  \int_0^\infty~{{\rm d}k\over k}\ P_\kappa(k)
  \left[J_1(k\theta)\right]^2,
\label{eqn:tophatvariance}
\end{equation}
where $J_1(x)$ is the first Bessel function of the first kind and $P_\kappa$
is the convergence power spectrum which depends on the source redshift
distribution $n(w)$ via
\begin{eqnarray}
  P_\kappa(k) &=& {9\over 4}\Omega_m^2
    \int_0^{w_H} {{\rm d}w \over a^2(w)}\ P_{3D}\left({k\over f_K(w)};w\right)
    \times\nonumber \\
  && \left[ \int_w^{w_H}{\rm d}w'\ n(w')\,{f_K(w'-w)\over f_K(w')}\right]^2,
\label{eqn:pofkappa}
\end{eqnarray}
where $w$ is the radial distance at redshift $z$, $f_K(w)$ is the
comoving angular distance to redshift $z$, and $\Omega_m$ is
the matter density parameter.  Thus $P_\kappa$ is the
weighted integral of the three-dimensional mass power spectrum,
$P_{3D}$, with a weight depending on $n(w)$. We imagine that $n(w)$ is
determined from a calibration sample with an error $\delta n(w)$.
The shear covariance matrix will then depend explicitly on a term
of the form $\langle \delta n(w) \delta n(w')\rangle$, which has
off-diagonal power due to large-scale structure. It is difficult
to proceed analytically, especially if we wish to analyze the
distribution of fluctuations in $n(w)$. Instead we take a
numerical approach and simulate $n(w)$ by populating the dark matter
halos from N-body simulations with mock galaxies. A simplified model will
however tell us how a redshift uncertainty is likely to affect the
analysis of cosmic shear data. It was shown in \cite{BVWM97} that,
to first order in the perturbation regime,
for a power law power spectrum the top-hat shear variance at scale $\theta$
behaves like:
\begin{equation}
\langle \gamma^2\rangle \propto \sigma_8^2 ~ z_s^{1.7} ~
\Omega_m^{1.7} ~ \theta^{\left(n-1\over 2\right)},
\label{shearvar}
\end{equation}
where $z_s$ is the mean source redshift and $n$ and $\sigma_8$ are
the slope and amplitude of the matter power spectrum,
respectively. The mean redshift is degenerate with $\sigma_8$ and
$\Omega_m$, therefore, uncertainty in $z_s$ should act as an
unknown normalization constant, as we shall see below.

\subsection{Mock catalogs}  \label{sec:mock}

The basis of our mock catalogs is a large N-body simulation of a $\Lambda$CDM
cosmology.  The simulation used $512^3$ particles in a periodic cubical box
$256h^{-1}$Mpc on a side.  This represents a large enough cosmological volume
to ensure a fair sample of the Universe, while maintaining enough mass
resolution to identify galactic mass halos.  The cosmological model is chosen
to provide a reasonable fit to a wide range of observations with
$\Omega_{\rm m}=0.3$, $\Omega_\Lambda=0.7$,
$H_0=100\,h\,{\rm km}{\rm s}^{-1}{\rm Mpc}^{-1}$ with $h=0.7$,
$\Omega_{\rm B}h^2=0.02$, $n=0.95$ and $\sigma_8=0.9$.
The simulation was started at $z=50$ and evolved to the present with a TreePM
code \cite{TreePM}.  The full phase space distribution was dumped every
$128\,h^{-1}$Mpc from $z\simeq 2$ to $z=0$.  The gravitational force softening
was of a spline form, with a ``Plummer-equivalent'' softening length of
$18\,h^{-1}$kpc comoving.  The particle mass is $10^{10}h^{-1}M_\odot$ allowing
us to find bound halos with masses several times $10^{11}h^{-1}M_\odot$.

For each output we produced a halo catalog by running a ``friends-of-friends''
group finder (FoF; e.g.~\cite{DEFW}) with a linking length $b=0.15$ in units
of the mean inter-particle spacing.
This procedure partitions the particles into equivalence classes, by linking
together all particle pairs separated by less than a distance $b$.
This means that FoF halos are bounded by a surface of density roughly $140$
times the background density.  We use the sum of the masses of the particles
in the FoF group as our definition of the halo mass.

A past light cone was constructed by propagating a field, $4^\circ$ on a
side, at $2.5^\circ$ to one of the Cartesian axes of the box.
The periodicity of the simulation was used to extend the field beyond
$256\,h^{-1}$Mpc and early time outputs were used at further distances.
The halo information was transformed into the field coordinate system to
create a light cone halo distribution. This ensures that the halo distribution
is continuous, but does not repeatedly trace the same structure.
In all we traced 8 fields, spaced by $45^\circ$ in azimuth, down each of the
three principal axes of the box.  Since the same simulation was used to
create all of the fields they are not independent, but the differing
orientations and volumes probed in each field sample a wide range of
environments and projections.

Once the halo distribution is given we assigned galaxies using a simple halo
occupation distribution.  We assumed that each halo either contained a central
galaxy or did not, and if it contained a central galaxy it could also contain
a number of satellites.
The average number of galaxies in a halo of mass $M$ was
\begin{equation}
   \left\langle N_{\rm gal}(M)\right\rangle      \Theta(M-M_{\rm min})\left[ 1 + \left( {M-M_{\rm min}\over M_1}
\right) \right]
\end{equation}
where $\Theta$ is the Heaviside step function, $M_1=\mu M_{\rm
min}$ and we take $\mu=3$.  This form is a reasonable fit to the
observed HODs of magnitude limited samples of low redshift
galaxies (e.g.~\cite{SDSS}) if $\mu$ is chosen to be a little
higher than we have chosen it here. But both theoretical
\cite{ConWecKra} and observational \cite{YanWhiCoi} results
suggest $\mu$ is lower at $z\approx 1$, so we have chosen $\mu=3$
as a compromise.

For each field we divided the redshift interval $[0,2)$ into 15 bins and
adjusted the single remaining parameter, $M_1$, in the HOD in each bin to
ensure that $n(z)$ would match the form

\begin{equation}
  n(z)\,dz \propto \frac{\beta}{\Gamma\left(\frac{1+\alpha}{\beta}\right)}
    \left(\frac{z}{z_s}\right)^\alpha\;
    \exp\,\Big[{-\Big(\frac{z}{z_s}\Big)^\beta}\Big]\ {dz\over z_s}\ ,
\label{nofz}
\end{equation}
where $\alpha$, $\beta$ and $z_s$ are the free parameters and
$n(z)$ is normalized to unit area. The simulations are built with
$\alpha=\beta=2$. The absolute number counts are chosen to have
$\approx 15$ or 30 galaxies per square arcminute in the fields and
$z_0$ such that $\langle z\rangle=0.7$ or $1.0$.  The required
$M_1(z)$ were all smooth curves with a minimum just below
$10^{12}\,h^{-1}M_\odot$ near $z\approx 0.5$, being shallow to
low-$z$ and rising to several times $10^{12}\,h^{-1}M_\odot$ at
higher $z$. Once $M_1(z)$ was known the halos were populated with
galaxies assuming Poisson statistics for the number of satellites.
Galaxies were assumed to trace an NFW profile \cite{NFW} within
the halos, and redshift space distortions were included by
assuming galaxies faithfully trace the dark matter velocity field.
Our mock catalogues do not include a detailed description of
galaxy formation or merging, therefore it can only be a rough
description of the reality. In particular they do not contain any
information regarding the apparent magnitude, size and absolute
luminosity of the galaxies.

\subsection{Fitting $n(z)$}

For each of the two $n(z)$ models, we have twenty four independent
$4^\circ\times 4^\circ$ fields from which we construct a set of
redshift calibration catalogues (sub-fields) for various areas
ranging from 5 square arcminutes to 4 square degrees. The list of
calibration survey sizes is $S=[5.3,14,56,225,900,3600,14400]$
square arcmins (where the last two are 1 and 4 square degrees
respectively). For each calibration survey we measure the redshift
distribution $n(z)$ from the galaxy distribution in that sub-field
and fit it with the three parameter function given by Eq.
\ref{nofz}. For each calibration sample size, a limited number of
samples can be tiled in one 16 square degree mock field. For
instance there are only four 4 square degree samples per field,
but you can cut 3600 samples with 16 square arcmin each. We
compute the number count covariance matrix between redshift bins
for each calibration sample size $S$ and average the result over
twenty four independent realizations. The distribution of the
parameters $\alpha$, $\beta$ and $z_s$ are stored, for the are
used later to calculate the shear covariance due to the redshift
distribution sampling variance. We arbitrarily choose to work with
$10$ redshift bins with a redshift spacing $\Delta z=0.23$.

\begin{figure}
\begin{center}
\resizebox{5.5in}{!}{\includegraphics{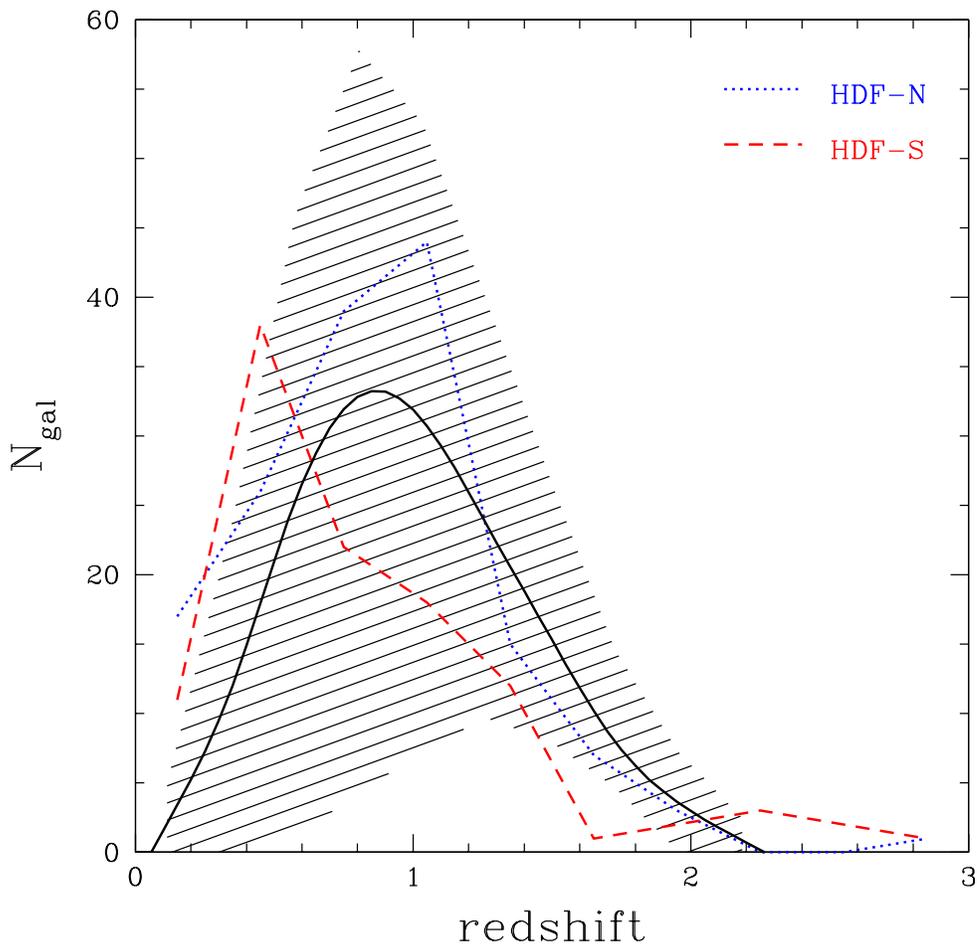}}
\end{center}
\caption{Redshift number counts in the Hubble Deep Fields North
and South (blue dotted and red dashed lines respectively) for a
magnitude cut $m_I=24.5$. The solid lines show the average number
counts from the high galaxy density mock catalogue in a
calibration survey of $5.3$ arcmin square (i.e. matching the HDF
area). The gray area show the measured r.m.s. in the high density
mock catalogue.} \label{fig:HDFcounts}
\end{figure}

The redshift distribution in each mock field follows the input
form, Eq.~(\ref{nofz}), quite well but large fluctuations, driven
by the spatial clustering of galaxies (large-scale structure), are
clearly visible.  As we average over the different mock fields
these fluctuations average away, but the field-to-field
fluctuations are far larger than the Poisson error in the counts
would predict \cite{Peebles}. Figure \ref{fig:HDFcounts} shows the
HDF North and South number counts (solid lines) for an AB
magnitude cut at $m_I=24.5$, which is the typical limiting
magnitude of most existing and many of the planned lensing
surveys. Each HDF field is $5.3$ arcmin square. This figure shows
that a $5.3$ arcmin square calibration survey from the mock
catalogue gives similar number counts to the HDFs. Given the large
error, resulting from the small HDF survey area, we conclude that
our mock catalogues provide a statistical description of reality
that is sufficient for the purposes of this paper.

\begin{figure}
\begin{center}
\resizebox{5.5in}{!}{\includegraphics{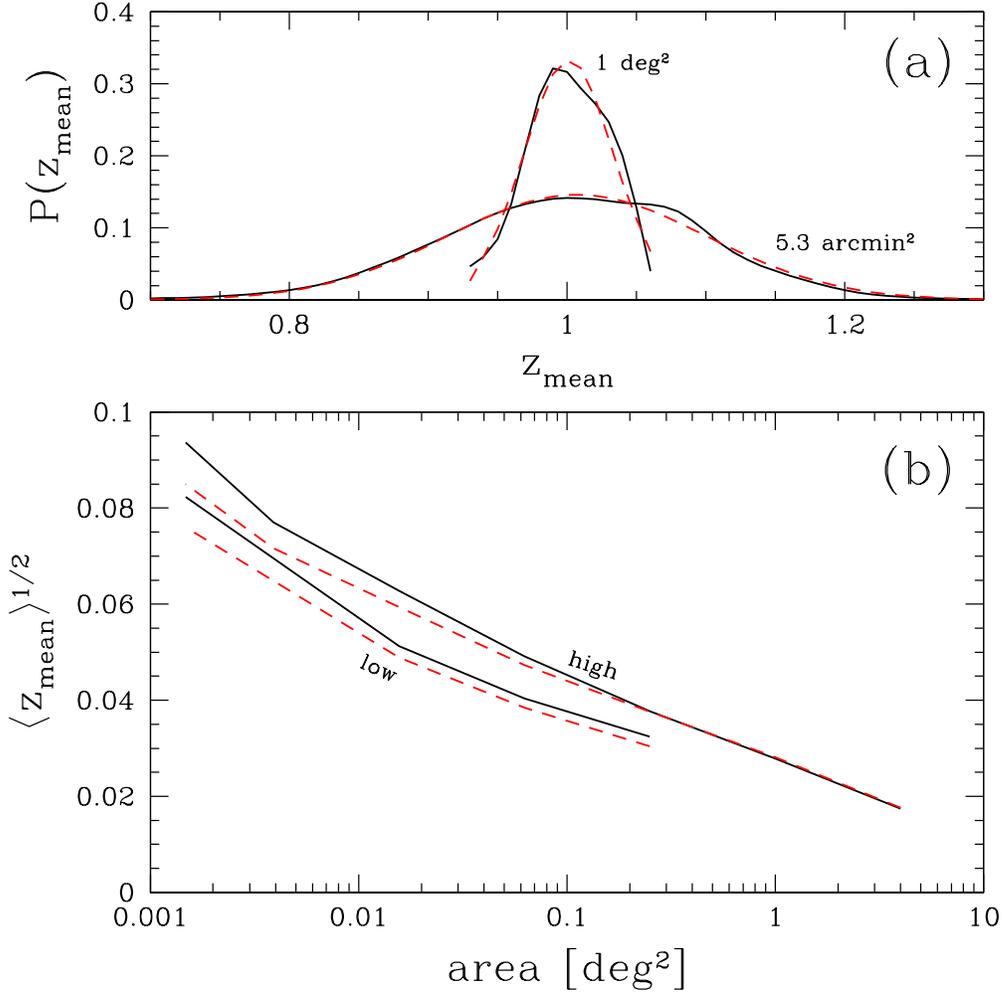}}
\end{center}
\caption{Top panel: the average source redshift obtained from al
the fitted $n(z)$ (dark solid lines). Red dashed lines show the
best Gaussian fit to the solid lines. The distribution of the
average source redshift distribution directly measured from the
mock catalogues is very similar to the dark solid lines. The
narrow distribution corresponds to a one square degree survey. The
broad distribution corresponds to a $5.3$ arcmin square (HDF)
survey area. Bottom panel: The error in the average source
redshift distribution as a function of the mock calibration sample
area. The top two lines are for the high redshift, high number
density, mock galaxy catalogue. The two bottom lines for the low
redshift low number density case. The dark solid lines are from
the fitted n(z) and the red dashed lines are directly measured in
the mock catalogues.} \label{fig:zmean_pdf}
\end{figure}

Top panel of Figure \ref{fig:zmean_pdf} shows the average source
redshift measured from the catalogues and from the fitted $n(z)$.
The dotted lines indicate that the average redshift is well
described by a Gaussian distribution, which we assume in the
analysis that follows.
Note that the average redshift (as shown in Figure 2) and the
variance (not shown) of the fit to n(z) are not affected by the
choice of the parametrization Eq.(\ref{nofz}).

\section{Results}  \label{sec:results}

\subsection{Redshift sample variance and co-variance}


A first estimation of the uncertainty introduced into cosmological
parameter estimation by the redshift sampling error is given by
the scatter of the measured mean redshift in the mock calibration
samples. The bottom panel of Figure \ref{fig:zmean_pdf} shows the
r.m.s.~of the mean redshift with respect to the true average
redshift for different calibration sample areas, assuming
contiguous survey coverage. The top two lines correspond to the
mock catalogue with mean redshift of $1.0$ and it shows the
scatter of the mean redshift for the fitted distribution (solid
line) and the distribution measured directly from the mock
catalogues (dashed line). The bottom two lines show the same but
for the lower redshift mock catalogue z=0.7. On average, the input
values $\alpha=2$ and $\beta=2$ are recovered, but there are
significant field-to-field variations. This comparison shows that
the solid and dashed lines almost overlap, meaning that the
fitting procedure does not introduce a significant excess of
scatter. It also shows that the sampling error is only slightly
changed between the two different mock catalogues. Interestingly,
we observe that a even a 4 square degree redshift sample gives a
mean redshift precision of $\sim 2$\%. This corresponds roughly to
an uncertainty in $\sigma_8$ at the same level (as illustrated by
Eq.\ref{shearvar}). It is also interesting to note that the
uncertainty of the mean redshift measured for a one square degree
field is in agreement to the dispersion measured in the
photometric redshift distribution in the four, one square degree,
CFHTLS deep fields \cite{IAM}. A detailed analysis of the redshift
sample requirements and the resulting effect on the measurement of
$\sigma_8$ is given in Sections 4 and 5.  It should be noted here,
however, that the largest {\it complete} spectroscopic redshift
calibration samples that will be available for the next few years
are limited to magnitude $m_I\simeq 24$ (VVDS and DEEP2 Groth
Strip) totaling $\sim 2.5$ square degrees. These redshift surveys
are clearly not big enough to calibrate the future lensing surveys
that will image hundreds of square degrees with the expectation of
achieving sub-percent accuracy on the measurement of cosmological
parameters.

Figure \ref{fig:zsource_rms_vs_area} demonstrates that it is
particularly inefficient to calibrate the redshift distribution
with a large contiguous redshift survey: the precision of 3\% on
the mean redshift with a one square degree survey is attainable
with only 10 independent HDF-sized redshift surveys (totaling
0.015 square degrees), because the error decreases as square root
of the number of independent fields. This conclusion agrees with
\cite{Ish} who found that only a small number of spectroscopic
redshifts are needed in order to calibrate a redshift
distribution: the authors found that only a thousand redshifts are
necessary to get the required redshift accuracy for lensing
studies on nearly the whole sky. This is true if we do not worry
about galaxy selection (from color, type, morphology, etc...),
meaning that the calibration sub-field could be as small as we
want (reduced to a single object as stated in \cite{Ish}). We find
that this statement is still valid even when non-linear source
clustering is included, which clearly demonstrates that the number
of independent calibration fields is much more important than the
size of the calibration fields themselves.

\begin{figure}
\begin{center}
\resizebox{5.5in}{!}{\includegraphics{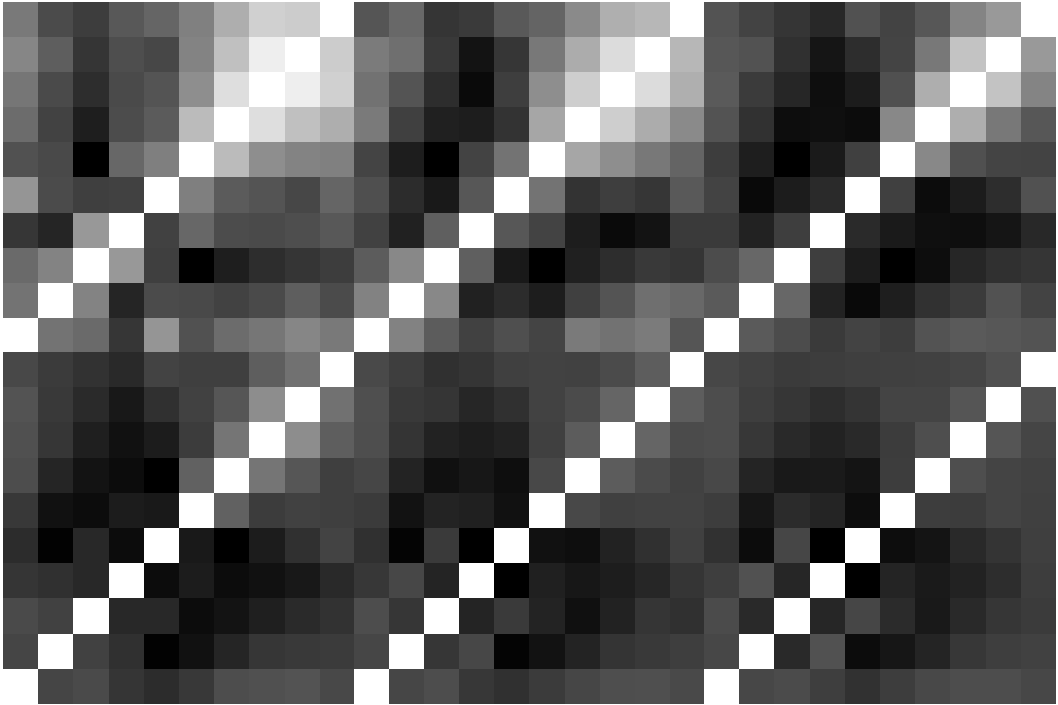}}
\end{center}
\caption{Redshift distribution covariance matrices for different
calibration survey size.  From top-left to bottom-right (reading
ordering) the survey size is 4 sq.deg., 1 sq.deg., 0.25 sq.deg.,
225 sq.arcmin, 56 sq.arcmin and 16 sq.arcmin. For small area
calibration samples, the covariance matrix becomes diagonal
although the error is larger than Poisson statistics would imply
(see Figure \protect\ref{fig:sample_poisson_variance}).}
\label{fig:rcoeffcombine}
\end{figure}

\begin{figure}
\begin{center}
\resizebox{5.5in}{!}{\includegraphics{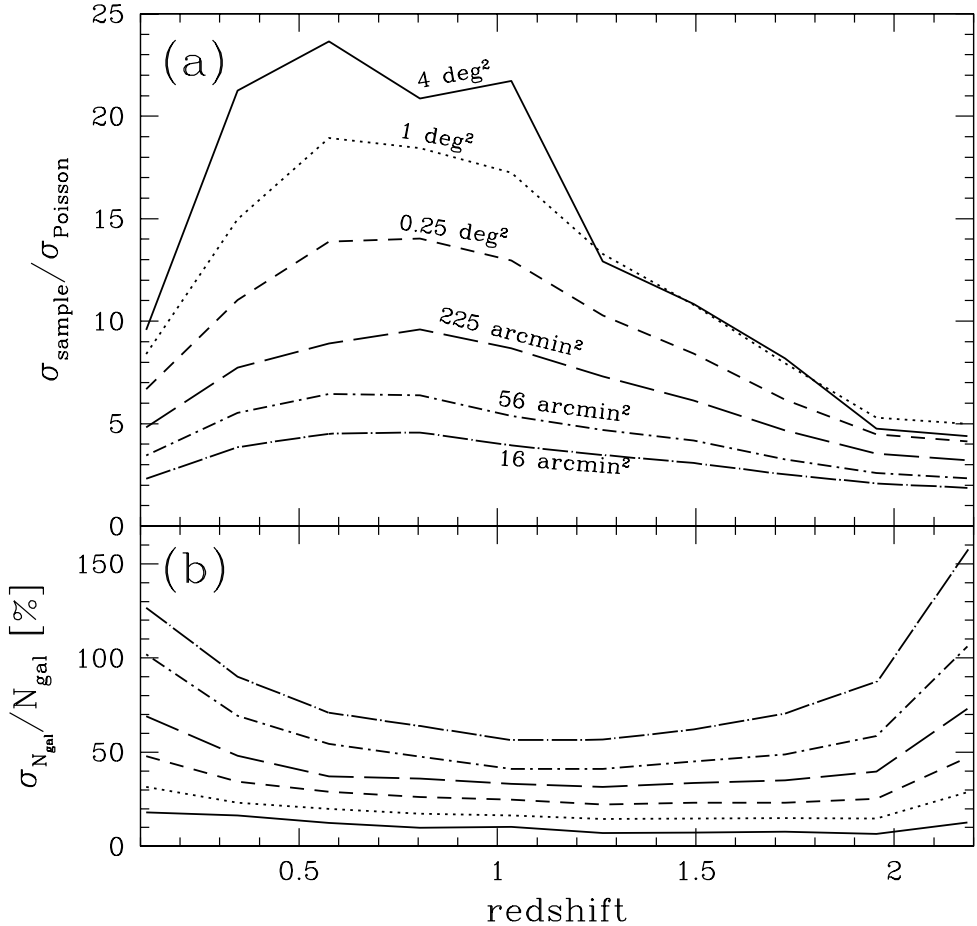}}
\end{center}
\caption{Top panel: the solid lines show the ratio between the
sample variance r.m.s. and the Poisson error r.m.s. From top to
bottom, the lines correspond to the calibration surveys of size 4
sq.deg., 1 sq.deg., 0.25 sq.deg., 225 sq.arcmin, 56 sq.arcmin and
16 sq.arcmin respectively. Bottom panel: lines show the fractional
sampling error (sample r.m.s. counts over the counts) for the same
six calibration surveys where the bottom dashed line corresponds
to the 4 sq.deg. calibration survey, and the top dashed line
corresponds to the 16 sq.arcmin calibration survey.}
\label{fig:sample_poisson_variance}
\end{figure}

Figure \ref{fig:rcoeffcombine} shows the covariance matrix of the
galaxy counts in ten redshift bins between $0<z<2.3$ for six
different calibration sample sizes.  Small sample sizes show
little correlation between bins, but the fluctuation amplitude is
much larger than Poisson statistics would imply. This is shown in
Figure \ref{fig:sample_poisson_variance} which plots the observed
variance to the Poisson prediction. Even for a 16 square arcmin
redshift sample the noise is 5 times the Poisson expectation! This
largely explains the significant difference between the
redshift number counts
of the two HDFs. The decline of the sample variance r.m.s over
Poisson error at large redshifts is as result of probing more
uncorrelated structures as the survey volume grows.

\subsection{The Impact on Cosmic Shear Analysis}

In this section, we directly investigate the effect of the
redshift sample variance on cosmological parameters measurements.
We focus on the constraints on $\sigma_8$ and $\Omega_m$, since
they are the main parameters that weak lensing is sensitive to
(see Eq.\ref{shearvar}). As described in Section 2.3, for each
mock catalogue, we have many sub-catalogues of different sizes. We
have a measure of the values of $\alpha$, $\beta$ and $z_s$ in
each of the sub-catalogues which we use to compute the covariance
matrix of the shear top-hat variance $\langle\gamma^2\rangle$,
holding the mass power spectrum $P(k)$ fixed to its theoretical
value. The result is averaged over the different mock catalogue
realizations. In this way, we directly obtain the contribution of
the redshift sample variance to the shear covariance matrix ${\bf
C_z}$, in other words, this corresponds to an increased error in
the shear measurement due to the n(z) sampling error. The full
shear covariance matrix is then given by ${\bf C}={\bf C_s}+{\bf
C_n}+{\bf C_z}$, where ${\bf C_s}$ is the cosmic variance
(calculated according to \cite{SvWKM}, which does include
non-linear amplitude of the power spectrum, but not the
non-Gaussian statistics) and ${\bf C_n}$ is the statistical noise.
A maximum likelihood calculation of the parameters $\sigma_8$ and
$\Omega_m$ is performed, assuming a fiducial model with
$\Omega_m=0.3$, $\Omega_\Lambda=0.7$, $\sigma_8=0.9$ and a power
spectrum shape parameter $\Gamma=0.21$. The fiducial source
redshift distribution is given by Eq. \ref{nofz}. The likelihood
function is given by
\begin{equation}
L\propto {\rm Exp}\left( -{1\over 2}d^{\rm T}~{\bf
C}^{-1}~d\right),
\end{equation}
where $d=\langle \gamma^2\rangle-\langle \gamma^2\rangle_{\rm
fiducial}$ is the measured shear top-hat variance as function of
scale minus the fiducial model top-hat variance. $d$ is given in
the scale range $[0.4,140]$ arcmin, which covers the scales of
interest where the effect of redshift distribution sampling
variance is important.

\begin{figure}
\begin{center}
\resizebox{5.5in}{!}{\includegraphics{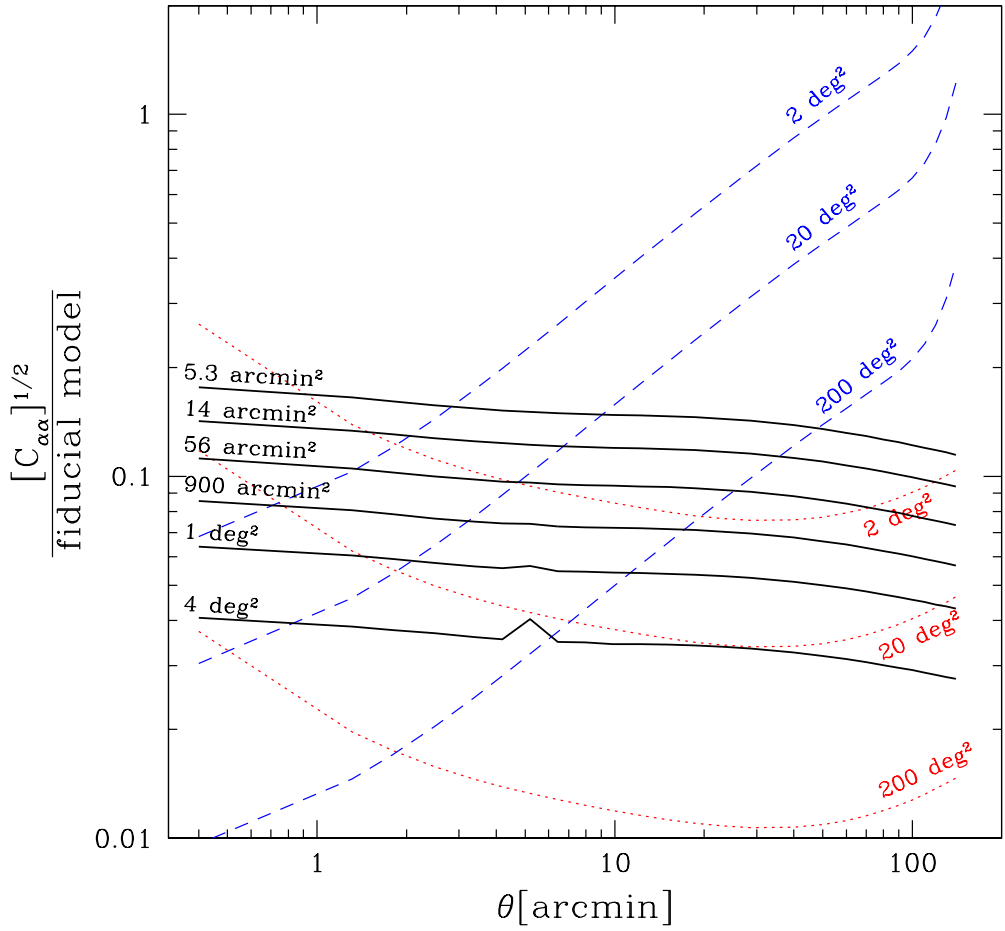}}
\end{center}
\caption{Different error contributions to the total lensing
covariance matrix as function of the smoothing scale $\theta$.
Each contribution is plotted as square root of the diagonal
elements of the matrix over the signal of the fiducial model. Red
dotted lines and blue dashed lines show the statistical noise and
cosmic variance (along the diagonal of the covariance matrix)
respectively. From top to bottom, the lines correspond to lensing
surveys of size 2 sq.deg, 20 sq.deg. and 200 sq.deg. The intrinsic
ellipticity is chosen to be $\sigma_e=0.36$ with the number
density of galaxies set to 20 galaxies per arcmin square.  The
dark solid lines show the diagonal covariance matrix due to source
redshift sample variance. From top to bottom, the redshift
calibration sample is 5.3, 14, 56, 900 arc-minutes, 1 and 4 square
degrees respectively.} \label{fig:covariancematrices}
\end{figure}

The behavior of the redshift distribution sampling variance is
particularly interesting as, to first approximation, it behaves
like an unknown normalization constant in the power spectrum, as
expected from Eq.\ref{shearvar} for a single source redshift. For
a broad redshift distribution, this is rather unexpected, however,
as different realizations of redshift distribution can vary
greatly for different lines-of-sight, changing not only the mean
source redshift but the entire shape of the distribution as well.
We find that the main contribution of the redshift sample variance
can be characterized by an effective mean source redshift, as if
the sources were located in a single redshift plane. Figure
\ref{fig:covariancematrices} shows that this is not the case to
second order, where the r.m.s. of the ratio between the diagonal
elements of the sampling variance covariance matrix and the
fiducial model shows a slight dependence on the smoothing scale.
The off-diagonal components, not shown here, are such that the
correlation coefficient over the entire matrix is $\sim 1$ within
$2\%$ accuracy. Therefore, the redshift uncertainty is mostly
degenerate with $\sigma_8$ and with a shear calibration error,
which is defined as the factor $1\pm\epsilon$ between the observed
$\gamma_{\rm obs}$ and true shear $\gamma_{\rm true}$, such that
$\gamma_{\rm obs}=(1\pm\epsilon)\gamma_{\rm true}$ . A shear
calibration error\footnote{Note that a $\epsilon $ calibration
error corresponds to a $(1\pm \epsilon)^2\simeq 1\pm 2\epsilon$
error in the power spectrum or shear variance.} can arise from a
galaxy shape measurement error which is quantified in
\cite{STEP1}. With this knowledge, it is then easy to include the
redshift sampling variance caused by non-linear large scale
structures in parameter forecasts by simply increasing the
uncertainty in the shear calibration factor. Another interesting
feature of Figure \ref{fig:covariancematrices} is that the n(z)
sample variance has the largest impact in the scale range $[1,10]$
arc-minutes (although it largely depends on the survey type),
where the sum of the two main sources of error in cosmic shear
surveys reaches a minimum. Above a few tens arcminutes, the cosmic
variance dominates, below one arcminute, the shot noise becomes
dominant (note that in this figure the intrinsic ellipticity
distribution is chosen to be $\sigma_e=0.36$ with the number
density of galaxies set to 20 galaxies per arcmin square). The
optimal redshift calibration survey would then be designed such
that its contribution to the shear covariance never exceeds the
value of the crossing point between the statistical and cosmic
variance errors. Changing the survey characteristic numbers will
change the redshift calibration requirements. A detailed
discussion on survey design including the redshift calibration
issue is included in the next Section.

\section{Designing future lensing surveys}

\subsection{Statistical noise versus cosmic variance}
\begin{figure}
\begin{center}
\resizebox{5.5in}{!}{\includegraphics{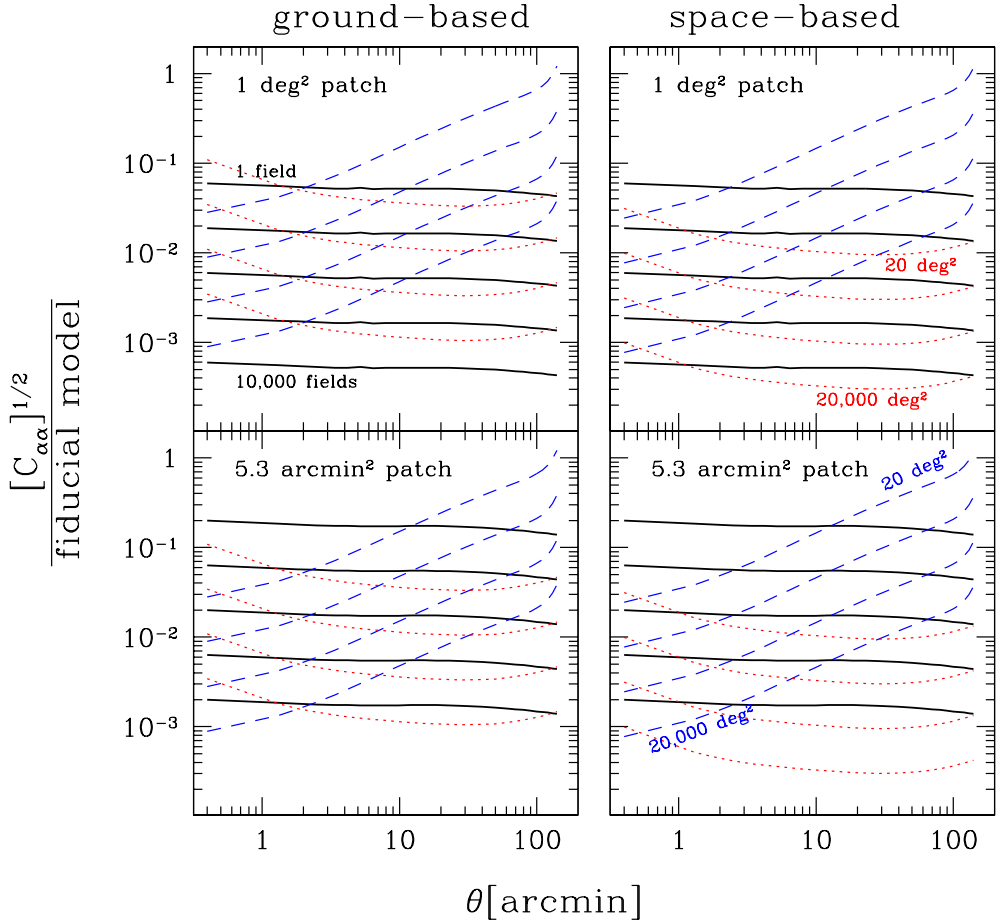}}
\end{center}
\caption{The different sources of noise for ground-based (left
panels) and space-based (right panels) lensing surveys, normalized
to the fiducial model. The dark solid lines show the diagonal of
the redshift sampling covariance matrix, from top to bottom they
correspond to a calibration sample with $1$, $10$, $100$, $1000$,
$10000$ fields, assuming a sparse survey, each field is a one
sq.deg. patch (top panels) and $5.3$ arcmin square patch (bottom
panels). The blue dashed lines show the cosmic variance and the
red dotted lines the statistical noise for four lensing surveys
sizes. From top to bottom the lines correspond to lensing surveys
of $20$, $200$, $2000$, $20000$ sq.deg.. The ground based survey
(left panels) assumes a galaxy number density $n_{gal}=15$ per
arcmin square and shape noise $\sigma_e=0.44$, while the shallow
space based survey (right panels) assumes $n_{gal}=35$ and
$\sigma_e=0.36$.} \label{fig:Cmatricesgroundspace}
\end{figure}

We now turn to forecasting to make predictions regarding the
minimal requirements for redshift calibration surveys. As it was
mentioned previously, we want the redshift sample variance noise
to be, at most, equal to the crossing point of the cosmic variance
and statistical noise errors. This constraint sets the size of the
needed calibration survey for a fixed set of lensing survey
characteristics. The best approach is to make a sparse calibration
survey (see Section 3.1) in order to minimize the sample variance
between the different fields. Therefore we adopt the following
strategy: we assume that the redshift calibration survey is made
of a collection of $N_{\rm cal}$ uncorrelated redshift
(spectrometric or photometric) surveys. The size of the individual
patches is either one square degree or $5.3$ square arcmin (HDF
size), from which we compare the performance. The error on the
mean redshift scales as $\sqrt{N_{\rm cal}}^{-1}$, with a redshift
uncertainty of nearly $3\%$ for the one sq.deg. patch and $9\%$
for the $5.3$ arcmin square (see Figure \ref{fig:zmean_pdf}). We
consider four lensing surveys of size $(20, 200, 2000, 20000)$
square degrees, and five redshift calibration surveys of $(1, 10,
100, 1000, 10000)$ fields (sparsely sampled with patches of one
square degree or 5.3 square arc-minutes each). We also consider a
ground-based type of statistical noise, with a number density of
galaxies per arcmin square $n_{gal}=15$ and an ellipticity noise
$\sigma_e=0.44$, and a shallow space-based type of statistical
noise with $n_{gal}=35$ and $\sigma_e=0.36$ (consistent with DUNE
\cite{ref}).

Figure \ref{fig:Cmatricesgroundspace} compares the different noise
amplitudes for the different surveys. It is interesting to note
the difference between ground and space based surveys. If we
assume one square degree calibration patches, then a $20000$
sq.deg. ground based survey can be calibrated with a $1000$ square
degree redshift survey, however this is clearly not good enough
for the $20000$ square degrees space based survey, which will not
perform significantly better than a $2000$ square degrees survey
if the calibration sample is not increased to $10000$ sq.deg. This
discussion is particularly relevant if we want to measure the
cosmic shear power spectrum in the $1-10$ arcmin range. One could
imagine a ground based $10$ sq.deg. calibration redshift survey,
for example, which would already be very time consuming if the
goal was to match space-based lensing data in terms of depth and
galaxy number density. The redshift sampling errors resulting from
the small size of the calibration survey would effectively wash
out the lensing signal producing results similar to a $200$
sq.deg. lensing survey for scales below $2$ arcmin and $2000$
sq.deg. for scales between $2$ and $10$ arcmin. If the calibration
patch is only $5.3$ arcmin square (HDF area), the same redshift
accuracy is reached for patches of one square degrees with only
ten times more independent fields, even though the HDF is nearly
$700$ times smaller than one square degree.

This analysis also demonstrates that one cannot separate the issue
of redshift calibration and shear calibration. Both need to be
below the sum of the statistical and cosmic variance errors in
order for the lensing survey to be fully efficient. As we can see
from Figure \ref{fig:Cmatricesgroundspace}, a shear calibration
error of $1\%$ would dramatically limit the cosmological
information that can be extracted from a lensing survey larger
than 200 square degrees, and any improvement in redshift
calibration below that limit would indeed be a waste of observing
time, unless the shear calibration is also improved. This is
consistent with \cite{Ish} who discuss why it is useless to
improve either redshift error or the shear calibration. Figure
\ref{fig:Cmatricesgroundspace} shows that both have to be below
the sum of statistical and cosmic variance errors if we want the
lensing survey to deliver its full potential. It is worth noting
that this discussion also applies to the {\it self calibration}
regime proposed by \cite{HTBJ} in which the shear calibration is
treated as a free parameter included with the cosmological
parameters we want to measure. Self calibration could however
allow us, by combining second and third order shear statistics, to
reach higher calibration precision than the shape measurement
accuracy.

The conclusion of this section is that there is a tight relation
between the source redshift sampling variance, the shear
calibration error and the statistical noise of a lensing survey,
the later being determined by the survey characteristics. This
relation is important for the design of lensing surveys. For
instance if the combined redshift and shear calibration errors are
at the $1\%$ level, then a $20000$ square degrees space based
lensing survey with $35$ galaxies per arcmin square would do as
well as if the number density of galaxies was only $0.35$ per
arcmin square.

\subsection{Calibration requirements for future lensing surveys}

In the previous section, we demonstrated that the accuracy of the
shear calibration sets strong limits on the ability of lensing
surveys to probe small angular scales $\theta<30$ arcmin. Although
the shear calibration error has been included in cosmological parameter
forecasts as an unavoidable source of systematics \cite{HTBJ}, to
date, there has been no discussion on how shear calibration could
limit the design of a lensing survey.
In this section we therefore try to answer this
question by comparing a large sample of observing strategies, and
derive the shear calibration requirement for each of them.
Note that we will refer generically to `shear
calibration' when talking about redshift or shear
calibration errors, since, to first approximation, they impose the
same limitation to a lensing survey.  Our
analysis will result in a tight relation between the shear
calibration accuracy and the best lensing survey that one can
undertake, beyond which accumulating more data would not improve
anything unless the shear calibration is itself improved.

To model the redshift distribution $n(z)$ of different magnitude
limited surveys (i.e with $m<m_{\rm lim}$) we use the method
described in \cite{Heymans05}, such that

\begin{equation}
n(z)\left[m < m_{\rm lim} \right] = { \sum_{i=1}^{\rm lim} N(i) n(z,m_i)
\over \sum_{i=1}^{\rm lim} N(i) }
\end{equation}
where $n(z,m_i)$ is the redshift distribution of galaxies in
magnitude slice of width $\Delta m = 0.5$ with a maximum magnitude
$m_i$, and
$N (i)$ is the number density of resolved galaxies in each
magnitude slice. We model $n(z,m_i)$ using equation(5) with
$\alpha = 2.2$ and $\beta = 1.0$, corresponding to the
best-fitting shape parameters to the HDF photometric redshift
distribution from \cite{Budavari}.  Using these parameters $z_0=
z_m/2.87$ where the median redshift $z_m$ can be estimated from
the redshift-magnitude relation of \cite{Heymans05} ($z_m = -3.132
+ 0.164 m$ where the AB magnitude $m$ is  for the F606W HST
filter). We estimate the number density of resolved galaxies in
each magnitude slice $N(i)$ from the galaxy number counts in the
HST ultra-deep field \footnote{\it HST UDF:
www.stsci.edu/hst/udf}, considering two different cases; a
ground-based survey with 0.7 arcsec seeing and a deep space-based
survey (e.g. SNAP) where the resolution is limited by 0.1 arcsec
pixels. We define a source to be adequately resolved for lensing
studies if the object's half light radius is greater than the
resolution limit set by the atmospheric seeing (ground), or pixel
scale (space). Table \ref{nofztable} lists the resulting number
density of sources and median source redshift for ground and
space-based surveys with different limiting magnitudes. One should
be careful here as these numbers were obtained from a small
field-of-view and are therefore sensitive to the sampling variance
discussed in this paper, this explains why the brightest magnitude counts
appear deeper (in redshift) from the ground than from space.
However, what is important for our purpose here, is the relative
evolution of the statistical noise and cosmic variance as function
of redshift.

\begin{figure}
\begin{center}
\resizebox{5.5in}{!}{\includegraphics{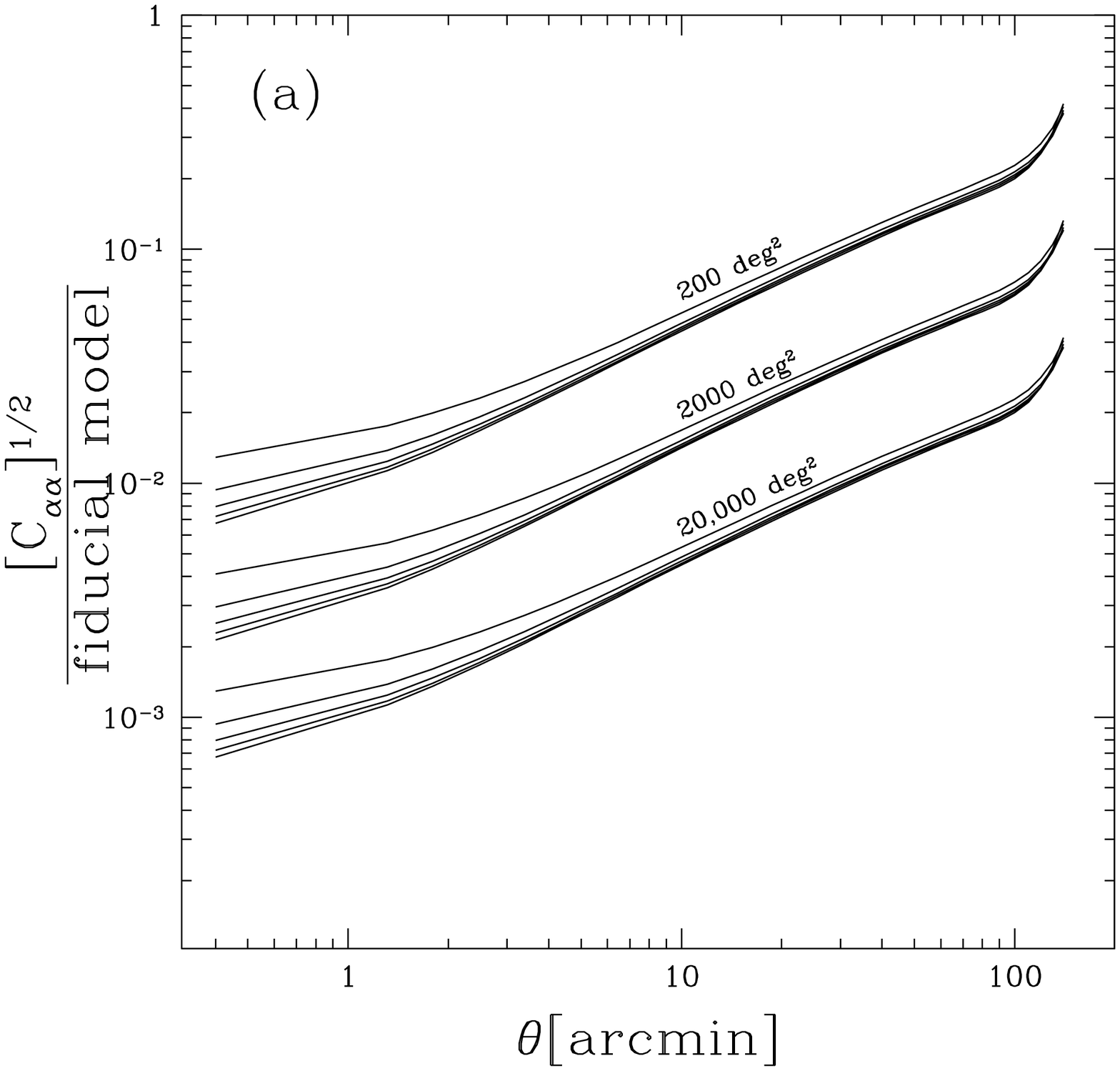}\includegraphics{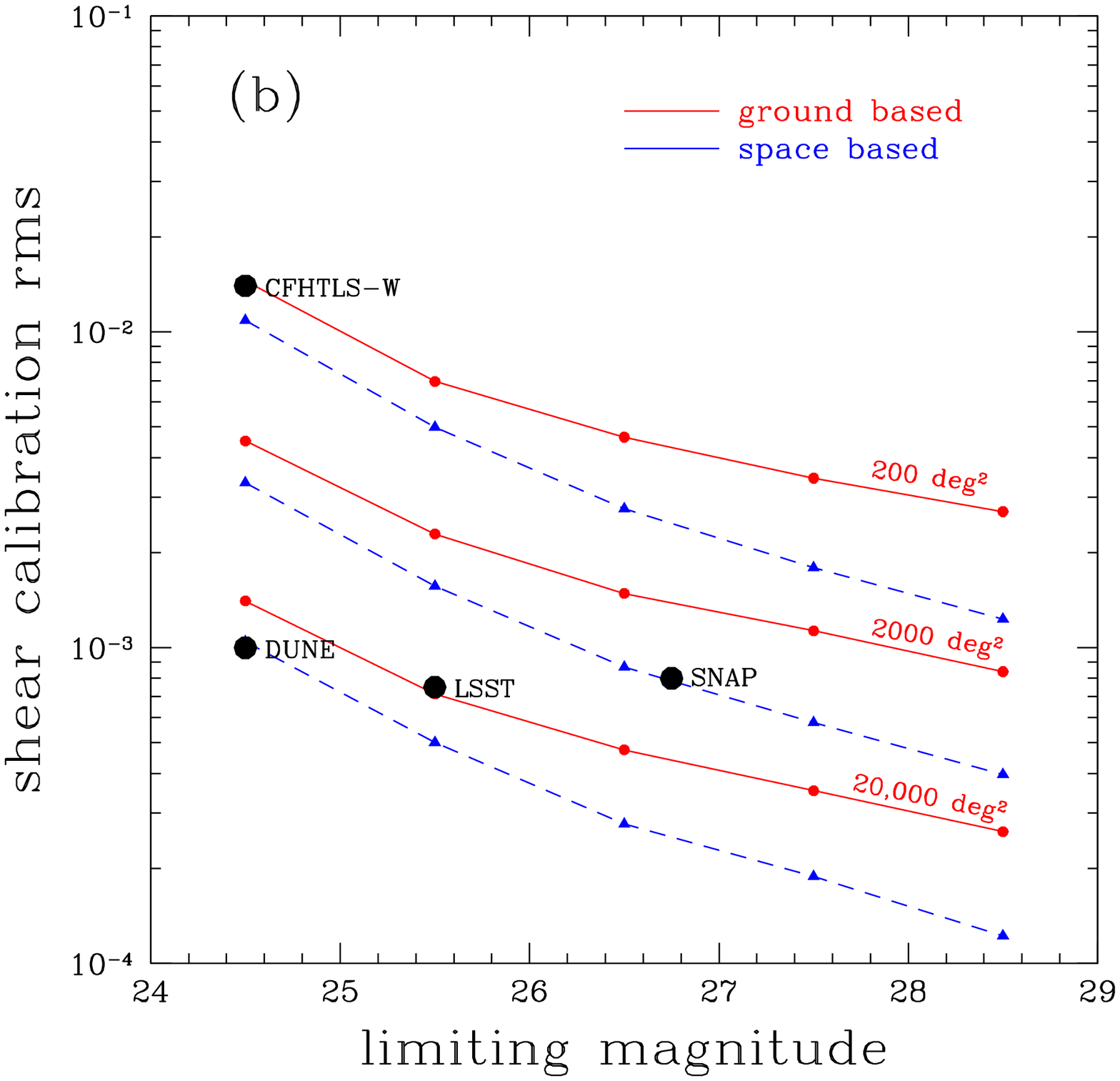}}
\end{center}
\caption{Left panel: the cosmic variance contribution for
different survey sizes. There are three curve bundles, from top to
bottom the curves correspond to survey sizes $200$, $2000$ and
$20000$ square degrees. In each bundle, the lines correspond to
different limiting magnitudes: 24.5, 25.5, 26.5, 27.5, 28.5. Right
panel: each line shows the calibration requirement for different
survey strategies, calculated as explained in Section 4.2. The
plot shows the optimal calibration r.m.s. for ground and space
based surveys with different depths. The dark bullets display some
the future and present weak lensing surveys.}
\label{fig:calibrequirement}
\end{figure}

Using this model, we calculate the covariance and
statistical noise matrices for ground and space
based surveys, with limiting magnitudes between $m_{\rm \lim} =24.5$ and
$m_{\rm \lim} = 28.5$. The left panel of
Figure \ref{fig:calibrequirement} shows that the covariance matrix
is relatively insensitive to the survey limiting magnitude:
surveys with very different $m_{lim}$ will have very different
lensing signal amplitude, but the covariance matrix will scale
accordingly. The covariance roughly scales as the inverse of the
survey area. This means that the efficiency of the different limiting
magnitude surveys will essentially differ by the change in the
statistical noise contribution to the covariance matrix, the ratio of the
cosmic variance to the lensing signal remaining the same.

With the data
from Table \ref{nofztable} we calculate the top-hat shear variance
for different models, and define the shear calibration requirement
as the particular point where the diagonal elements of the cosmic
variance and statistical noise matrices divided by the signal
cross each other. The right panel of Figure
\ref{fig:calibrequirement} shows the shear requirement as function
of the limiting magnitude and angular scale of the lensing survey
plotted along with some of the planned lensing surveys
(LSST\footnote{LSST, {\it www.lsst.org}}, DUNE {\cite{ref}} and
SNAP\footnote{SNAP, {\it snap.lbl.gov}}).

We fit the shear calibration requirement to the magnitude and survey size
from Figure \ref{fig:calibrequirement} and find
that the shear calibration $\epsilon$ (see Section 3.2) must be
below $\epsilon = \epsilon_0 10^{\beta (m-24.5)} \left(\Theta\over
200\right)^{-1/2}$, where $(\epsilon_0,\beta)=(0.015,-0.18)$ and
$(\epsilon_0,\beta)=(0.011,-0.23)$ for a ground and space based
survey respectively. The $\Theta$ dependence of this limit is
driven by the cosmic variance only (consistent with \cite{HTBJ}),
while the magnitude dependence is driven by the statistical noise
via the number counts.

\begin{table}
\caption{Table showing the average source redshift and galaxy number
density for
different limiting magnitude surveys where the calculation is described
in the text.
The survey limiting
F606W AB magnitudes are given in the first column,  the
average source redshift in the second,
and the last column is the number density
of galaxies per arc-minute square. The quoted numbers correspond
to a ground (space) based survey.} \label{nofztable}
\begin{center}
\begin{tabular}{|c|c|c|}
\hline
Survey depth & $\langle z_{source}\rangle$ & $n_{gal}$ \\
\hline
24.5  &      0.825 (0.787)  & 8 (13)\\
25.0  &      0.879 (0.869)  & 11 (20)\\
25.5  &      0.951 (0.948)  & 16 (30)\\
26.0  &      1.019 (1.044)  & 21 (45)\\
26.5  &      1.076 (1.128)  & 26 (65)\\
27.0  &      1.143 (1.216)  & 32 (91)\\
27.5  &      1.196 (1.285)  & 38 (124)\\
28.0  &      1.248 (1.358)  & 44 (166)\\
28.5  &      1.307 (1.440)  & 51 (222)\\
 \hline
\end{tabular}
\end{center}
\end{table}

\section{A note regarding previous cosmic shear measurements}

Given that the impact of redshift sampling variance has so far
been neglected, and that nearly all present lensing surveys rely
on the HDFs to calibrate their redshift distribution, one should
investigate whether the errors on published cosmic shear
measurements have been correctly estimated. Table
\ref{sigma8error} shows the error on the normalization of the
power spectrum $\sigma_8$ for different surveys where the redshift
calibration sample consists of two independent HDF-sized surveys.
One can see that for a lensing survey such as VIRMOS or
CFHTLS-WIDE in its present stage, although the redshift
uncertainty is large, the difference between assuming a Poisson
redshift distribution or the full sample covariance is only at the
$\sim 10$ percent level. This is not true for deeper surveys of
the same size (e.g model (20,1.0) in table~\ref{sigma8error}),
where the difference can be as large as $100$\%, as well as for
the complete 200 square degree CFHTLS-WIDE survey. {\it The
multi-color data of the CFHTLS will be crucial in order to produce
complete photometric redshift catalogues and thus achieve the
cosmic variance limited accuracy}. According to Figure
\ref{fig:calibrequirement}, one also needs to achieve a shear
calibration accuracy of one percent which is the current state of
the art of weak shear measurement \cite{STEP1}.

The recent third year WMAP release \cite{WMAP3} has shown that
$\sigma_8$ measured from CFHTLS \cite{HH06,ES06} is rather on the
high side. The redshift sample variance discussed in this work
could easily account for this difference: the HDFs have been
chosen as empty fields, and it is not unreasonable to believe that
their redshift distribution could significantly differ from the
average distribution as a consequence of selection effects. One
should also note that COMBO-17 \cite{COMBO-17} found a relatively low
$\sigma_8$, fully consistent with \cite{WMAP3}, using accurate
redshift information. Fortunately, the CFHTLS survey will soon
deliver photometric redshifts for the DEEP \cite{IAM}
and WIDE surveys and it
will become possible to calibrate the redshift distribution to the
accuracy required by the size of the lensing data set. Future work will
include a preliminary check of the CFHTLS results with the CFHTLS
DEEP photometric redshifts \cite{IAM} and a lensing analysis that
combines all surveys published to date, taking
into account a more realistic redshift distribution than the previously used
Hubble Deep Fields. In \cite{HH05}, it was mentioned that the RCS
photometric redshifts show that the actual mean redshift is larger
than the one from HDF. The RCS $\sigma_8$ should probably be about
$8\%$ lower. This is another good example why photometric
redshifts are essential.

\begin{table}
\caption{Table indicating the error on the measured $\sigma_8$
with two independent HDF sized samples as the redshift calibration
survey. The second column assumes Poisson error for the redshift
distribution plus cosmic variance.  The third column is for the
full redshift sample variance and cosmic variance, and the last
column assumes cosmic variance only. The first column specifies
the lensing survey type, where the first number indicates the area
in square degrees and the second number indicates the mean source
redshift. The number density of galaxies is always $15$ galaxies
per arc-minute square, which represents an average number count
for sources between redshift 0.7 and 1., and the shape noise is
$0.44$, these numbers correspond to a typical ground based survey
for these redshift depth.} \label{sigma8error}
\begin{center}
\begin{tabular}{|c|c|c|c|}
\hline
model & $\Delta_{Poiss} \sigma_8$ & $\Delta_{All} \sigma_8$ & $\Delta_{Covar} \sigma_8$ \\
\hline
(4,0.7) & 0.136 & 0.143 & 0.098 \\
(20,0.7) & 0.084 & 0.092 & 0.037   \\
(200,0.7) & 0.051 & 0.063 & 0.013  \\
(4,1.0) & 0.058 & 0.077 & 0.056  \\
(20,1.0) & 0.030 & 0.057 & 0.025   \\
(200,1.0) & 0.018 & 0.046 & 0.008  \\
 \hline
\end{tabular}
\end{center}
\end{table}

\section{Conclusions}  \label{sec:conclusions}

We have studied the effect of redshift sample variance on cosmic
shear measurements using a realistic distribution of galaxies
embedded in dark matter halos. This source of error occurs when
photometric redshifts cannot be obtained for the entire lensing
survey, which is the case for all current surveys, and will be the
case for some of the future surveys that will be unable to follow
up in multi-color all of the fainter galaxies that will be used in
the lensing analysis. We derived what minimum requirements a
redshift calibration sample should have in order to make the
redshift distribution error negligible compared to the statistical
and cosmic variance errors. We have shown that the redshift sample
variance behaves like a shear calibration factor to first
approximation, even for the general case when galaxies are
distributed within large scale structures. We have shown that even
when non-linear source clustering is included, the best redshift
sampling strategy is still a sparse sample. However it is clear
that the best way to avoid redshift sampling issues is to have a
complete photometric redshift survey, which is also required to
remove contamination from intrinsic galaxy alignments (see for
example \cite{HH03}) and shear-ellipticity correlations
\cite{King05,CH06}.

The shear and redshift calibrations are both important for
designing future lensing surveys. An optimal use of lensing survey
time is to guarantee that the statistical and cosmic variance
errors are not smaller than the summed redshift and shear
calibration errors. This is particularly critical for small
angular scales (less than a few tens arcminutes), and it puts
strong constraints on the useful maximum galaxy number density a
lensing survey should have. We have derived the calibration
requirements using realistic galaxy number counts from the Hubble
Ultra-Deep Field.

Among the work that remains is to investigate how the photometric
redshift errors couple to the sampling variance error. A more
realistic analysis will also include realistic galaxy populations
with color, morphology and size distributions. Our analysis will
also have to be extended to include tomography studies. Some recent papers
discuss the effect of imperfect photometric measurement on
cosmic shear studies in the tomographic case \cite{HTBJ,MHH}. The
authors allow for different error models in the estimated mean
redshift and they concluded that the mean redshift must be known
to great accuracy, a few $10^{-3}$, which is similar to our calibration
requirement for almost full sky surveys.

Future lensing surveys are designed such that the number density
of source galaxies is higher than the current value of $\sim 20$
galaxies per arc-minute square. That means they have the ambitious
goal of measuring the mass power spectrum at a relative precision
of $10^{-3}$. This is clearly possible only if both the source
redshift and the shear calibration errors are known to a similar
accuracy, a level of precision which is still far below the actual
state of the art of weak shear measurement \cite{STEP1}.

\bigskip

We thank Mustapha Ishak-Boushaki for discussions on the topics
developed in this work, Eric Linder, Dragan Huterer and Gary
Bernstein for useful comments on the manuscript and Alexandre
R\'efr\'egier for discussions regarding the DUNE project. We are
very grateful to Tamas Budavai for sharing his HDF photometric
redshift catalogues and the full probability distributions for
each object.  The simulations used here were performed on the
IBM-SP at NERSC. MJW was supported in part by NASA and the NSF.
This work uses data from the Hubble Ultra Deep Field (UDF) which
is a public HST survey made possible by Cycle 12 STSci Director's
Discretionary Time, programme GO/DD-9978.   LVW and HH are
supported by the Natural Sciences and Engineering Research
   Council (NSERC), the Canadian Institute for Advanced Research
   (CIAR) and the Canadian Foundation for Innovation (CFI),
CH is supported by a CITA National Fellowship, NSERC and CIAR.

\end{document}